\RequirePackage{ifpdf}
\ifpdf 
\documentclass[pdftex]{sigma}
\else
\documentclass{sigma}
\fi

\begin{document}
\allowdisplaybreaks

\renewcommand{\PaperNumber}{015}

\FirstPageHeading

\ShortArticleName{Second Order Superintegrable Systems in Three Dimensions}

\ArticleName{Second Order Superintegrable Systems\\ in Three Dimensions}

\Author{Willard MILLER}                                    
\AuthorNameForHeading{W. Miller}

\Address{School of Mathematics, University of Minnesota,
 Minneapolis, Minnesota, 55455, USA} 
 
\Email{\href{mailto:miller@ima.umn.edu}{miller@ima.umn.edu}}

\URLaddress{\href{http://www.ima.umn.edu/~miller/}{http://www.ima.umn.edu/\~{}miller/}}

\ArticleDates{Received October 28, 2005; Published online November 13, 2005}

\Abstract{A classical (or quantum) superintegrable system  on an 
$n$-dimensional Riemannian manifold  is an integrable  Hamiltonian system with potential that admits $2n-1$ 
functionally independent constants of the motion that are 
polynomial in the momenta, the maximum number  possible. 
If these constants of the motion are all quadratic, the system is second order superintegrable.  
Such systems have remarkable properties. Typical properties are that  
1)~they are integrable in multiple ways and comparison of ways of integration leads to new facts about the systems,  
2)~they are multiseparable, 3)~the second order symmetries generate a
closed quadratic algebra and in the quantum case the representation
theory of the quadratic algebra yields important facts about the
spectral resolution of the Schr\"odinger operator and the other
symmetry operators,    and 4)~there are deep
connections with expansion formulas relating classes of special
functions and with the theory of Exact and Quasi-exactly Solvable systems. 
For $n=2$ the author,  E.G.~Kalnins and  J.~Kress,
have worked out the structure of these systems and classified all of the possible spaces and potentials.  
Here I discuss our recent work and announce new results for the much more difficult case $n=3$. We
consider classical superintegrable systems with nondegenerate
potentials in three dimensions and on a conformally flat real or complex space. 
We show that  there exists a standard structure for such systems, based on the algebra
of $3\times 3$ symmetric matrices, 
and that the quadratic algebra always closes at order~6. We describe the 
St\"ackel transformation, an invertible  conformal mapping between 
superintegrable structures on distinct spaces, and give evidence indicating that all
our superintegrable systems are St\"ackel transforms 
of systems on complex Euclidean space or the complex 3-sphere. 
We also indicate how to 
extend the classical 2D and 3D superintegrability 
theory to include the operator (quantum) case.}

\Keywords{superintegrability; quadratic algebra; conformally flat spaces}

\Classification{37K10; 35Q40; 37J35; 70H06; 81R12}

\section{Introduction and examples}  
In this paper I will report on
recent  and ongoing work with E.G.~Kalnins and J.~Kress to uncover
the structure of second order superintegrable systems, both classical
and quantum mechanical. I~will concentrate on the basic ideas; 
the details of the proofs can be found elsewhere. The results on the
quadratic algebra structure of 3D conformally flat systems with
nondegenerate potential have appeared recently. The results on the 3D St\"ackel transform and 
multiseparability of superintegrable systems with nondegenerate 
potentials are announced here. 

Superintegrable systems can lay claim to be the most symmetric solvable systems in mathematics though, technically, 
many such systems admit no group symmetry.  In this paper I will only consider superintegrable systems on complex conformally flat spaces. This is no restriction at all in two dimensions.
An
$n$-dimensional complex Riemannian space is conformally  flat if and only if
it admits a set of local coordinates $x_1,\dots,x_n$ such that the
contravariant metric tensor takes the form
$g^{ij}=\delta^{ij}/\lambda(\boldsymbol{x})$. Thus the metric is
$ds^2=\lambda(\boldsymbol{x})\big(\sum\limits_{i=1}^ndx_i^2\big)$.  A classical
superintegrable system ${\cal H} =\sum\limits_{ij}g^{ij}p_ip_j+V(\boldsymbol{x})$ on
the phase space of this manifold   is one that 
admits $2n-1$ functionally independent  generalized
symmetries (or constants of the motion) 
${\cal S}_k$, $k=1,\dots,2n-1$ with ${\cal S}_1={\cal H}$ where the ${\cal S}_k$  
are polynomials in the momenta $p_j$.  That is, $\{{\cal H},{\cal S}_k\}=0$ where 
\[ 
\{f,g\}=\sum_{j=1}^n(\partial_{x_j}f\partial_{p_j}g-\partial_{p_j}f\partial_{x_j}g)
\]
is the Poisson bracket for functions $f(\boldsymbol{x},\boldsymbol{p})$,
$g(\boldsymbol{x},\boldsymbol{p})$ on phase space \cite{WOJ,EVA,EVAN,FMSUW,FSUW,MSVW,CALO,CIMC}.  
It is easy to see that $2n-1$ is the maximum possible 
number of functionally independent symmetries and, locally, 
such (in general nonpolynomial) symmetries always exist.   
The system is second order superintegrable if the $2n-1$ functionally independent symmetries can be chosen 
to be quadratic in the momenta. Usually a superintegrable system is
also required to be integrable, i.e., it is assumed that $n$ 
of the constants of the motion are in involution, 
although I will not make that assumption in this paper. 
Sophisticated tools such as $R$-matrix theory can be applied 
to the general study of superintegrable systems, e.g., 
\cite{Skly1989, FT1987, Har2000}. However, the most detailed and 
complete results are known for second order superintegrable systems because 
separation of variables methods for the associated Hamilton--Jacobi 
equations can be applied. Standard orthogonal separation of variables techniques are associated
with second-order symmetries, e.g., \cite{EIS1949, MIL,KMJ80,MIL83, ERNIE, MIL88} 
and multiseparable Hamiltonian systems provide numerous examples of 
superintegrability. Thus here I concentrate on second-order
superintegrable systems,  on those in which the symmetries 
take the form ${\cal S}=\sum a^{ij}(\boldsymbol{x})p_ip_j+W(\boldsymbol{x})$,
quadratic in the momenta.

There is an analogous definition for second-order quantum
superintegrable systems with Schr\"odinger operator 
\[ H=\Delta +V(\boldsymbol{x}),\qquad
\Delta=\frac{1}{\sqrt{g}}\sum_{ij}\partial_{x_i}\big(\sqrt{g}g^{ij}\big)\partial_{x_j},
\]
  the Laplace--Beltrami operator  plus a
potential function~\cite{EIS1949}. Here there are $2n-1$
second-order  symmetry operators 
\[ S_k=\frac{1}{\sqrt{g}}\sum_{ij}\partial_{x_i}\big(\sqrt{g}a^{ij}_{(k)}\big)\partial_{x_j}+W^{(k)}(\boldsymbol{x}),
\qquad k=1,\dots,2n-1
\]
 with $S_1=H$
and $[H,S_k]\equiv HS_k-S_kH=0$. Again multiseparable systems yield
many examples of superintegrability, though 
 not all multiseparable systems are superintegrable and not 
 all second-order superintegrable systems are multiseparable. 

The basic motivation for studying superintegrable systems 
is that they can be  solved explicitly and in multiple ways. 
It is the information gleaned from comparing the distinct solutions and 
expressing one 
solution set in terms of another that is a primary reason for their interest. 

Two dimensional second order superintegrable systems have been studied
and classified by the author and his collaborators in a recent series
of papers \cite{KKM20041, KKM20042,KKM20051,KKM20052}.  
Here we concentrate on three dimensional (3D) systems where new complications arise. We
start with  some  simple 3D examples   to illustrate some of 
the main features of superintegrable systems. 
(To make clearer the connection with quantum theory and Hilbert 
space methods we shall, for these examples alone, adopt standard physical  normalizations, such as using the factor
$-\frac12$ in front of the free Hamiltonian.)
Consider the 
Schr\"odinger  equation $H\Psi=E\Psi$ or 
 ($\hbar=m=1$, $x_1=x$, $x_2=y$, $x_3=z$)
\[
{ H}\Psi 
= - {\frac12}
\left(\frac{\partial^2}{\partial x^2} +
\frac{\partial^2}{\partial y^2}+ \frac{\partial^2}{\partial z^2}
\right)\Psi
+ V(x,y,z) \Psi = E\Psi.
\]
The generalized anisotropic oscillator corresponds to
the 4-parameter  potential
\[
V(x,y,z) = \frac{\omega^2}{ 2}\left(x^2+y^2+4(z+\rho)^2\right)  +
\frac{1}{2}
\left[\frac{k^2_1-\frac{1}{4}}{x^2}+
\frac{k^2_2-\frac{1}{4}}{y^2}
\right].
\]
(This potential is ``nondegenerate'' in a precise sense that I will
explain later.)  
The corresponding Schr\"odinger equation has separable solutions
in five coordinate systems: Cartesian coordinates, cylindrical polar
coordinates, cylindrical elliptic coordinates, cylindrical parabolic
coordinates and parabolic coordinates. The energy eigenstates for this
equation are degenerate and important  special function identities
arise by expanding one basis of separable eigenfunctions in terms of
another. 
A  second order symmetry operator for this equation is a second order
linear diffe\-ren\-tial operator $S$ such that $[H,S]=0$, where
$[A,B]=AB-BA$.   A basis for these operators~is
\begin{gather*}
M_1=\partial ^2_x-\omega ^2x^2+ \frac{k^2_1-\frac14}{ x^2},
\qquad
M_2=\partial^2_y - \omega^2y^2 - \frac{k^2_2-\frac14}{ y^2},
\\
P=\partial^2_z-4\omega^2(z+\rho)^2,\qquad
L = L^2_{12} -
\left(k^2_1 - \frac14\right)\frac{y^2}{ x^2} -
\left(k^2_2-{\frac14}\right)\frac{x^2}{ y^2} - {\frac12},
\\
S_1 = - {\frac12}
(\partial_x L_{13} + L_{13} \partial_x)+\rho\partial^2_x + (z+\rho) \left(\omega^2 x^2
- \frac{k^2_1-\frac14}{x^2}\right),
\\
S_2 = - {\frac12}
(\partial_y L_{23} + L_{23} \partial_y) +\rho\partial^2_y+ (z+\rho) \left(\omega^2 y^2
- \frac{k^2_2-\frac14}{y^2}\right),
\end{gather*}
where  $L_{ij}=x_i\partial_{x_j}-x_j
\partial_{x_i}$. It can be verified that these symmetries generate a ``quadratic
algebra''
that closes at level six. Indeed, the nonzero commutators of the above basis
are
\[
[M_1,L] = [L,M_2]=Q, \qquad
[L,S_1] = [S_2,L] = B, \qquad
[M_i,S_i] = A_i, \qquad
[P,S_i] = - A_i.
\]
Nonzero commutators of the basis symmetries with $Q$ (4th order symmetries) are
expressible in terms of the second order symmetries:
\begin{gather*}
[M_i,Q]=[Q,M_2]=4\{M_1,M_2\}+16\omega ^2L,
\qquad
[S_1,Q]=[Q,S_2]=4\{M_1,M_2\},
\\
[L,Q]=4\{M_1,L\}-4\{M_2,L\}+16\left(1-k^2_1\right)M_1-16\left(1-k^2_2\right)M_2.
\end{gather*}
There are similar expressions for commutators with $B$ and the
$A_i$. Also the squares of $Q$, $B$, $A_i$ and products such as $\{Q,B\}$,
(all 6th order symmetries) are all expressible in terms of 2nd order
symmetries. Indeed 
\begin{gather*}
Q^2={\frac83}\{L,M_1,M_2\}
+8\omega ^2\{L,L\}-16\left(1-k^2_1\right)M^2_1-16\left(1-k^2_2\right)M^2_2
\\
\phantom{Q^2=}{}+\frac{64}{ 3}\{M_1,M_2\}-\frac{128}{3}\omega ^2L
-128\omega^2\left(1-k^2_1\right)\left(1-k^2_2\right),
\\
\{Q,B\}=-{\frac83}\{M_2,L,S_1\}-
{\frac83}\{M_1,L,S_2\}+16\left(1-k^2_1\right)\{M_2,S_2\}+16\left(1-k^2_2\right)\{M_1,S_1\}
\\
\phantom{\{Q,B\}=}{}
-\frac{64}{ 3}\{M_1,S_2\}-\frac{64}{ 3}\{M_2,S_1\}.
\end{gather*}
Here $\{C_1,\dots,C_j\}$ is the completely symmetrized product of
operators $C_1,\dots,C_j$. (For complete details see \cite{KMP99}.) The point is that the algebra generated by
products and commutators of the 2nd order symmetries closes at order
6. This is a remarkable fact, and ordinarily not the case for an integrable system.

A counterexample to the existence of a quadratic algebra  in Euclidean space is given by the Schr\"odinger equation
with  3-parameter
extended Kepler--Coulomb 
potential: 
\[
\left(\frac{\partial^2 \Psi}{\partial x^2}
+ \frac{\partial^2 \Psi}{\partial y^2}
+ \frac{\partial^2 \Psi}{\partial z^2}\right) +
\left[2E + \frac{2\alpha }{ \sqrt{x^2+y^2+z^2}} -
\left(\frac{k^2_1-\frac14}{ x^2}+\frac{k^2_2 - \frac14}{ y^2}
\right)\right] \Psi = 0.
\]
This equation admits separable solutions in the four coordinates
systems: spherical, sphero-conical, prolate spheroidal and
parabolic coordinates. Again the bound states are degenerate   and important  
special function identities
arise by expanding one basis of separable eigenfunctions in terms of
another. However, the space of second order symmetries is only 5
dimensional and, although there are useful identities among 
the generators and commutators that enable one to derive spectral properties algebraically,  
there is no finite quadratic algebra structure. The
key difference with our first example is, as we shall show later,  that the 3-parameter Kepler--Coulomb
potential is degenerate and  it cannot be extended to a 4-parameter potential.

In \cite{KKM20051, KKM20052} there are examples of superintegrable 
systems on the 3-sphere that admit a~quadra\-tic algebra structure. 
A more general set of examples arises from a  space with metric
\[ ds^2=\lambda(A,B,C,D,\boldsymbol{x})\left(dx^2+dy^2+dz^2\right),
\]
 where
\begin{gather*}
\lambda=A(x+iy)+B\left(\frac34(x+iy)^2+\frac{z}{4}\right)+
C\left((x+iy)^3+\frac{1}{16}(x-iy)+\frac{3z}{4}(x+iy)\right)\\
\phantom{\lambda=}{}
+D\left(\frac{5}{16}(x+iy)^4+\frac{z^2}{16}+\frac{1}{16}\left(x^2+y^2\right)+\frac{3z}{8}(x+iy)^2\right).
\end{gather*}
The nondegenerate classical potential  is 
$V=\lambda(\alpha,\beta,\gamma,\delta,\boldsymbol{x})/\lambda(A,B,C,D,\boldsymbol{x})$. 
If $A=B=C=D=0$ this is a nondegenerate metric on complex 
Euclidean space. The quadratic algebra always closes, and for general values of $A$,
$B$, $C$, $D$  the space is not of constant curvature.  As will be apparent later.
This is an example of a superintegrable system  that is
St\"ackel equivalent to a~system on complex Euclidean space. 

Observed common features of superintegrable systems are that they are
usually multiseparable and that 
the eigenfunctions of one separable system can be expanded in terms of
the eigenfunctions of another. This is the source of 
nontrivial special function expansion theorems~\cite{KMJP}. 
The symmetry operators are in formal self-adjoint form and 
suitable for spectral analysis. Also, the quadratic algebra identities 
allow us to relate eigenbases and eigenvalues of one symmetry operator 
to those of another. The representation theory of the abstract quadratic algebra 
can be used to derive spectral properties of the second order generators 
in a manner analogous to the use of Lie algebra representation theory to derive spectral
properties of quantum systems that admit Lie symmetry algebras,
\cite{ KMJP,BDK,CDas, SPS}. (Note however that for superintegrable 
systems with nondegenerate potential, there is no first order 
Lie symmetry.)  

Another  common feature of quantum superintegrable systems is that
they can be modified by a gauge transformation so that  the Schr\"odinger and
symmetry operators are acting on a space of polynomials~\cite{KMT}. 
This is closely related to the theory of exactly and  quasi-exactly
solvable systems~\cite{USH,VILE}. The characterization 
of ODE quasi-exactly solvable systems as embedded in PDE superintegrable 
systems provides considerable insight into the nature of these
phenomena~\cite{KMP2005}.

The classical analogs of the above examples are obtained by the
replacements $\partial_{x_i}\to p_{x_i}$ and modification of the
potential by curvature terms. Commutators go over to Poisson brackets. 
The operator symmetries become second order constants of the motion. Symmetrized operators become products of functions. The quadratic algebra 
relations simplify: the highest order terms agree with the operator case
 but there are fewer nonzero lower order terms.

Many examples of 3D superintegrable  systems are known, although they
have not been classified~\cite{GPS, KKMP, KKW, KKMW, RAN, KMWP}. 
Here,  we employ a theoretical method based on integrability conditions to derive structure 
common to all such systems, with a view to complete classification, at
least for classical systems with nondegenerate potentials. We show 
that for  systems with nondegenerate potentials there exists a standard 
structure based on the algebra of $3\times3$ symmetric matrices,  
and that the quadratic algebra closes at level~6. For 2D nondegenerate 
superintegrable systems we earlier showed that  the $3=2(2)-1$ functionally
independent constants of the motion were (with one exception) also
linearly independent, so at each regular point we could find a unique
constant of the motion that matches a quadratic expression in the
momenta at that point. However, for 3D systems we have only $5=2(3)-1$
functionally independent constants of the motion and the quadratic forms
span a 6 dimensional space. This is a major problem. However, for
nondegenerate potentials we  prove the ``5 implies 6 Theorem''
to show that the space of second order constants of the motion is in
fact 6 dimensional: there is a symmetry that is functionally dependent
on the symmetries that arise from superintegrability, but linearly
independent of them. With that result established, the treatment of
the 3D case can proceed  in analogy with the nondegenerate 2D case treated
in \cite{KKM20041}.  Though the details are quite complicated,  
the spaces of truly 2nd, 3rd, 4th and 6th  order constants of the motion can be 
shown to be  of dimension 6, 4, 21 and 56,   
respectively and  we can construct explicit bases for the 4th and 
6th order constants in terms of products of the 2nd order constants. 
This means that there is a quadratic algebra structure.

Using this structure we can show that all 3D superintegrable systems with 
nondegenerate potential are multiseparable. We  study  the St\"ackel transform, or coupling constant 
metamorphosis~\cite{BKM, HGDR}, for 3D classical superintegrable systems.
This is a conformal transformation of 
a superintegrable system on one space to a superintegrable 
system on another space. We give evidence  that all nondegenerate 3D superintegrable 
systems are St\"ackel transforms  of constant curvature systems, just as 
in the 2D case, though we don't completely settle the issue.  This provides the theoretical 
basis for a complete classification of 3D superintegrable 
systems with nondegenerate potential, a program that is underway. 
Finally we indicate  the quantum analogs of our results for  3D classical systems.

\section{Conformally flat spaces in three dimensions}

We  assume that there is a coordinate
 system $x$, $y$, $z$ and a nonzero function $\lambda(x,y,z)=\exp G(x,y,z)$ such that the
 Hamiltonian is
\[
{\cal H}=\frac{p_1^2+p_2^2+p_3^2}{\lambda}+V(x,y,z).
\]
A quadratic constant of the motion  (or generalized symmetry)
\[
{\cal S}=\sum ^3_{k,j=1}a^{kj}(x,y,z)p_kp_j+W(x,y,z)\equiv
{\cal L}+W,\qquad a^{jk}=a^{kj} 
\]
must satisfy   
$
\{{\cal H},{\cal S}\}=0$,
i.e.,
\begin{gather*}
a_i^{ii}=-G_1a^{1i}-G_2a^{2i}-G_3a^{3i},\nonumber\\
2a^{ij}_i+ a_j^{ii}=-G_1a^{1j}-G_2a^{2j}-G_3a^{3j},\qquad i\ne j,\nonumber\\
 a^{ij}_k+a^{ki}_j+a^{jk}_i=0,\qquad i,j,k\ {\rm distinct}
\end{gather*}
and
\begin{gather}
W_k=\lambda\sum_{s=1}^3a^{sk}
V_s,\qquad k=1,2,3.\label{potc}
\end{gather}
(Here a subscript $j$ denotes differentiation with respect to $x_j$.)
The requirement that $\partial_{x_\ell} W_j=\partial_{x_j}
W_\ell$, $\ell\ne j$ leads from
(\ref{potc}) to the
second order Bertrand--Darboux  partial differential equations for the potential.
\begin{gather}\label{BertrandDarboux}
\sum_{s=1}^3\left[V_{sj}\lambda a^{s\ell}-V_{s\ell}\lambda a^{sj}+
V_s\left(
(\lambda a^{s\ell})_j-(\lambda
a^{sj})_\ell\right)\right]=0.
\end{gather}

  For second order superintegrabilty in 3D there must
be five functionally independent constants of the motion 
(including the Hamiltonian itself). Thus  the Hamilton--Jacobi 
equation admits four   additional constants of the 
motion: 
\[ 
{\cal S}_h=\sum_{j,k=1}^3a^{jk}_{(h)}p_kp_j+W_{(h)}={\cal L}_h+W_{(h)},\qquad h=1,\dots,4.
\]
We assume that the four functions ${ {\cal S}}_h$ together with ${\cal H}$ are functionally
independent in the six-dimensional phase space. 
(Here the possible $V$  will always be assumed to 
form a vector space and we require functional independence 
for each such $V$ and the associated $W^{(h)}$. 
This means that we require that the five quadratic forms ${\cal L}_h$, ${\cal H}_0$ are 
functionally independent.)  In \cite{KKM20051} it is shown 
that the matrix of the 15 Bertrand--Darboux equations for the potential  
has rank at least 5, hence we can solve for the second derivatives of the potential in the form
\begin{gather}
V_{22}=V_{11}+A^{22}V_1+B^{22}V_2+C^{22}V_3,\nonumber\\
V_{33}=V_{11}+A^{33}V_1+B^{33}V_2+C^{33}V_3,\nonumber\\
V_{12}= \phantom{V_{11}+{}}A^{12}V_1+B^{12}V_2+C^{12}V_3,\nonumber\\
V_{13}= \phantom{V_{11}+{}}A^{13}V_1+B^{13}V_2+C^{13}V_3,\nonumber\\
V_{23}= \phantom{V_{11}+{}}A^{23}V_1+B^{23}V_2+C^{23}V_3.\label{veqn1a}
\end{gather}
If the matrix  has rank $>5$ then there will be
additional conditions of the form
$D^{1}_{(s)}V_1+D^{2}_{(s)}V_2+D^{3}_{(s)}V_3=0$. Here the $A^{ij}$,
$B^{ij}$, $C^{ij}$, $D^{i}_{(s)}$ are
functions of $x$ that can be calculated explicitly. 
For convenience we take $A^{ij}\equiv A^{ji}$, $B^{ij}\equiv B^{ji}$, $C^{ij}\equiv C^{ji}$.
 
Suppose now that the superintegrable system is such that the rank is exactly 5 so that the relations are only 
 (\ref{veqn1a}). Further, suppose the integrability conditions for system  (\ref{veqn1a})  are satisfied identically. 
In this case we say that the potential is {\it nondegenerate}. Otherwise the potential is {\it degenerate}.  
If $V$ is nondegenerate then at any point $\boldsymbol{x}_0$, where the $A^{ij}$, $B^{ij}$, $C^{ij}$ are 
defined and analytic, there is a unique solution $V(\boldsymbol{x})$ with
arbitrarily prescribed values of $V_1(\boldsymbol{x}_0)$, $V_2(\boldsymbol{x}_0)$,
$V_3(\boldsymbol{x}_0)$, $V_{11}(\boldsymbol{x}_0)$ (as well as the value of $V(\boldsymbol{x}_0)$ itself.)
The points $\boldsymbol{x}_0$ are called {\it regular}. 
The points of singularity for the $A^{ij}$, $B^{ij}$, $C^{ij}$ form a manifold of dimension $<3$.
Degenerate potentials depend on fewer parameters. For example,
it may be that the rank of the Bertrand--Darboux equations is exactly 
5 but the integrability conditions are not satisfied identically. 
This occurs for the generalized Kepler--Coulomb potential. 

From this point on we assume that $V$ is nondegenerate.
 Substituting the requirement for a~nondegenerate potential  (\ref{veqn1a}) into the
Bertrand--Darboux equations (\ref{BertrandDarboux}) we obtain three equations for the
derivatives $a^{jk}_i$, the first of which is
\begin{gather*}
\left(a^{11}_3-a^{31}_1\right)V_1+\left(a^{12}_3-a^{32}_1\right)V_2+(a^{13}_3-a^{33}_1)V_3\nonumber
\\
\qquad{}+a^{12}\left(A^{23}V_1+B^{23}V_2+C^{23}V_3\right)-\left(a^{33}-a^{11}\right)
\left(A^{13}V_1+B^{13}V_2+C^{13}V_3\right)\nonumber
\\
\qquad{}-a^{23}\left(A^{12}V_1+B^{12}V_2+C^{12}V_3\right)+a^{13}\left(A^{33}V_1+B^{33}V_2+C^{33}V_3\right)\nonumber
\\
\qquad{}=\left(-G_3a^{11}+G_1a^{13}\right)V_1+\left(-G_3a^{12}+G_1a^{23}\right)V_2+
\left(-G_3a^{13}+G_1a^{33}\right)V_3,\nonumber
\end{gather*}
and the other two are obtained in a similar fashion.

Since $V$ is a nondegenerate potential we can  equate coefficients of $V_1$,
$V_2$, $V_3$, $V_{11}$ on each side of the  conditions $\partial_1V_{23}=\partial_2V_{13}
=\partial_3V_{12}$, $\partial_3V_{23}=\partial_2V_{33}$, etc., to obtain
   integrability conditions, the simplest of which include
\begin{gather*}
A^{23}=B^{13}=C^{12},\qquad B^{12}-A^{22}=C^{13}-A^{33},\nonumber\\
B^{23}=A^{31}+C^{22},\qquad C^{23}=A^{12}+B^{33},\nonumber\\
A^{12}_1+B^{12}A^{12}+A^{33}_2+A^{33}A^{12}+B^{33}A^{22}+C^{33}A^{23}=A^{23}_3+B^{23}A^{23}+C^{23}A^{33},\nonumber\\
A^{13}_2+A^{13}A^{12}+B^{13}A^{22}+C^{13}A^{23}=A^{23}_1+B^{23}A^{12}+C^{23}A^{13}\nonumber\\
\qquad{} =A^{12}_3+A^{13}A^{12}+B^{12}A^{23}+C^{12}A^{33}.\nonumber
\end{gather*}

Using the nondegenerate potential condition and the Bertrand--Darboux equations  we can solve for
all of the first partial derivatives $a^{jk}_i$  of a  quadratic symmetry to obtain
\begin{gather} \label{symmetryeqnsc}
a^{11}_1=-G_1a^{11}-G_2a^{12}-G_3a^{13},\\
a^{22}_2=-G_1a^{12}-G_2a^{22}-G_3a^{23},\nonumber\\
a^{33}_3=-G_1a^{13}-G_2a^{23}-G_3a^{33},\nonumber\\
3a^{12}_1=a^{12}A^{22}-\left(a^{22}-a^{11}\right)A^{12}-a^{23}A^{13}+a^{13}A^{23}\nonumber\\
\phantom{3a^{12}_1=} {}+G_2a^{11}-2G_1a^{12}-G_2a^{22}-G_3a^{23},\nonumber\\
3a^{11}_2=-2a^{12}A^{22}+2\left(a^{22}-a^{11}\right)A^{12}+2a^{23}A^{13}-2a^{13}A^{23}\nonumber\\
\phantom{3a^{11}_2=}{} -2G_2a^{11}+G_1a^{12}-G_2a^{22}-G_3a^{23},\nonumber\\
3a^{13}_3=-a^{12}C^{23}+\left(a^{33}-a^{11}\right)C^{13}+a^{23}C^{12}-a^{13}C^{33}\nonumber\\
\phantom{3a^{13}_3=}{} -G_1a^{11}-G_2a^{12}-2G_3a^{13}+G_1a^{33},\nonumber\\
3a^{33}_1=2a^{12}C^{23}-2\left(a^{33}-a^{11}\right)C^{13}-2a^{23}C^{12}+2a^{13}C^{33}\nonumber\\
\phantom{3a^{33}_1=}{} -G_1a^{11}-G_2a^{12}+G_3a^{13}-2G_1a^{33},\nonumber\\
3a^{23}_2=a^{23}(B^{33}-B^{22})-\left(a^{33}-a^{22}\right)B^{23}-a^{13}B^{12}+a^{12}B^{13}\nonumber\\
\phantom{3a^{23}_2=}{} -G_1a^{13}-2G_2a^{23}-G_3a^{33}+G_3a^{22},\nonumber\\
3a^{22}_3=-2a^{23}\left(B^{33}-B^{22})+2(a^{33}-a^{22}\right)B^{23}+2a^{13}B^{12}-2a^{12}B^{13}\nonumber\\
\phantom{3a^{22}_3=}{} -G_1a^{13}+G_2a^{23}-G_3a^{33}-2G_3a^{22},\nonumber\\
3a^{13}_1=-a^{23}A^{12}+\left(a^{11}-a^{33}\right)A^{13}+a^{13}A^{33}+a^{12}A^{23}\nonumber\\
\phantom{3a^{13}_1=}{} -2G_1a^{13}-G_2a^{23}-G_3a^{33}+G_3a^{11},\nonumber\\
3a^{11}_3=2a^{23}A^{12}+2\left(a^{33}-a^{11}\right)A^{13}-2a^{13}A^{33}-2a^{12}A^{23}\nonumber\\
\phantom{3a^{11}_3=}{} +G_1a^{13}-G_2a^{23}-G_3a^{33}-2G_3a^{11},\nonumber\\
3a^{33}_2=-2a^{13}C^{12}+2\left(a^{22}-a^{33}\right)C^{23}+2a^{12}C^{13}-2a^{23}\left(C^{22}-C^{33}\right)\nonumber\\
\phantom{3a^{33}_2=}{} -G_1a^{12}-G_2a^{22}+G_3a^{23}-2G_2a^{33},\nonumber\\
3a^{23}_3=a^{13}C^{12}-\left(a^{22}-a^{33}\right)C^{23}-a^{12}C^{13}-a^{23}\left(C^{33}-C^{22}\right)\nonumber\\
\phantom{3a^{23}_3=}{} -G_1a^{12}-G_2a^{22}-2G_3a^{23}+G_2a^{33},\nonumber\\
3a^{12}_2=-a^{13}B^{23}+\left(a^{22}-a^{11}\right)B^{12}-a^{12}B^{22}+a^{23}B^{13}
\nonumber\\
\phantom{3a^{12}_2=}{} -G_1a^{11}-2G_2a^{12}-G_3a^{13}+G_1a^{22},\nonumber\\
3a^{22}_1=2a^{13}B^{23}-2\left(a^{22}-a^{11}\right)B^{12}+2a^{12}B^{22}-2a^{23}B^{13}\nonumber\\
\phantom{3a^{22}_1=}{} -G_1a^{11}+G_2a^{12}-G_3a^{13}-2G_1a^{22},\nonumber
\\
3a^{23}_1=a^{12}\left(B^{23}+C^{22}\right)+a^{11}\left(B^{13}+C^{12}\right)-a^{22}C^{12}-a^{33}B^{13}\nonumber\\
\phantom{3a^{23}_1=}{} +a^{13}\left(B^{33}+C^{23}\right)-a^{23}\left(C^{13}+B^{12}\right)
 -2G_1a^{23}+G_2a^{13}+G_3a^{12}.\nonumber\\
3a^{12}_3=a^{12}\left(-2B^{23}+C^{22}\right)+a^{11}\left(C^{12}-2B^{13}\right)-a^{22}C^{12}+2a^{33}B^{13}\nonumber\\
\phantom{3a^{12}_3=}{} +a^{13}\left(-2B^{33}+C^{23}\right)+a^{23}\left(-C^{13}+2B^{12}\right) 
-2G_3a^{12}+G_2a^{13}+G_1a^{23}.\nonumber\\
3a^{13}_2=a^{12}\left(B^{23}-2C^{22}\right)+a^{11}\left(B^{13}-2C^{12}\right)+2a^{22}C^{12}-a^{33}B^{13}\nonumber\\
\phantom{3a^{13}_2=}{} +a^{13}\left(B^{33}-2C^{23}\right)+a^{23}\left(2C^{13}-B^{12}\right)
-2G_2a^{13}+G_1a^{23}+G_3a^{12},\nonumber
\end{gather}
plus the linear relations
\begin{gather*}
 A^{23}=B^{13}=C^{12},\qquad B^{23}-A^{31}-C^{22}=0,\\
B^{12}-A^{22}+A^{33}-C^{13}=0,\qquad B^{33}+A^{12}-C^{23}=0.
\end{gather*}
Using the  linear relations we can express  $C^{12}$,
$C^{13}$, $C^{22}$, $C^{23}$ and $B^{13}$ in terms of the remaining  $10$ functions.

Since the above system of first order partial differential equations
is involutive the general
solution for the 6 functions $a^{jk}$ can depend on at most 6
parameters, the values $a^{jk}(\boldsymbol{x}_0)$ at a~fixed regular point
$\boldsymbol{x}_0$. For the  integrability conditions
 we define the vector-valued function
\[
\boldsymbol{h}(x,y,z)=\left(
  a^{11}\; a^{12}\; a^{13}\; a^{22}\; a^{23}\; a^{33}\right)
\]
and  directly compute the $6\times 6$ matrix functions ${\cal A}^{(j)}$ to get the first-order system
\[ 
\partial_{x_j}\boldsymbol{h}={\cal A}^{(j )}\boldsymbol{h},\qquad  j=1,2,3.
\]
The integrability conditions for this system are are
\begin{gather}\label{int5c}
{\cal A}^{(j)}_i\boldsymbol{h}-{\cal A}^{(i)}_j\boldsymbol{h}
={\cal A}^{(i)}{\cal A}^{(j)}\boldsymbol{h}-{\cal A}^{(j)}{\cal A}^{(i)}
\boldsymbol{h}\equiv [{\cal A}^{(i)},{\cal A}^{(j)}]\boldsymbol{h}.
\end{gather}
In terms of the $6\times 6$ matrices 
\begin{gather*}
 {\cal S}^{(1)}={\cal A}^{(3)}_2-{\cal A}^{(2)}_3- [{\cal
  A}^{(2)},{\cal A}^{(3)}],\qquad  {\cal S}^{(2)}={\cal A}^{(1)}_3-{\cal A}^{(3)}_1- [{\cal
  A}^{(3)},{\cal A}^{(1)}],
\\
 {\cal S}^{(3)}={\cal A}^{(2)}_1-{\cal A}^{(1)}_2- [{\cal
  A}^{(1)},{\cal A}^{(2)}],
\end{gather*}
the integrabilty conditions are 
\begin{gather}\label{int6c} {\cal S}^{(1)}\boldsymbol{h}={\cal S}^{(2)}\boldsymbol{h}={\cal
  S}^{(3)}\boldsymbol{h}=0.
\end{gather} 

\section[The $5 => 6$ Theorem]{The $\boldsymbol{5\Longrightarrow 6}$ Theorem}

Now assume that the system of equations (\ref{symmetryeqnsc}) admits a 6-parameter family of solutions
$a^{jk}$. (The requirement of superintegrability {\it appears} to guarantee
only a 5-parameter family of solutions.) Thus at any regular point we can prescribe the values of the
$a^{jk}$ arbitrarily. This means  that~(\ref{int5c}) or~(\ref{int6c}) holds
identically in $\boldsymbol{h}$. Thus ${\cal S}^{(1)}={\cal S}^{(2)}={\cal
  S}^{(3)}=0$.  This would be the analog of what happens in the 2D 
case where there are 3independent terms in the quadratic form and 
3~functionally (and linearly) independent symmetries. 
However, in the 3D case there are only 5~functionally independent symmetries, so we can't 
guarantee that the symmetry equations admit a 6-parameter family of solutions. Fortunately, by 
careful study of the integrability conditions of these equations and
use of the requirement that the potential is nondegenerate, we can prove the 
$5\Longrightarrow 6$ theorem~\cite{KKM20051}.

\begin{theorem}[$5\Longrightarrow 6$]  Let $V$ be a nondegenerate potential corresponding to
  a conformally flat space in $3$ dimensions that is superintegrable,
i.e., suppose $V$ satisfies the equations \eqref{veqn1a} whose
integrability conditions
 hold identically, and there are $5$ functionally
independent constants of the motion.  Then the space of second order symmetries
 for the Hamiltonian ${\cal H}=\left(p^2_x+p^2_y+p^2_z\right)/\lambda(x,y,z)+V(x,y,z)$
(excluding multiplication by a constant) is of
dimension $D= 6$. 
\end{theorem}

\begin{corollary} If ${\cal H} +V$ is a superintegrable conformally flat system with
nondegenerate potential, then the dimension of the space of $2$nd order
symmetries
\[
{\cal S}=\sum ^3_{k,j=1}a^{kj}(x,y,z)p_kp_j+W(x,y,z)
\]
is  $6$.  At any regular point $(x_0,y_0,z_0)$, and given constants
$\alpha^{kj}=\alpha^{jk}$, there is exactly one symmetry  ${\cal S}$ (up to an additive constant) such
that $a^{kj}(x_0,y_0,z_0)=\alpha^{kj}$. Given a set of $5$ functionally
independent $2$nd order  symmetries ${\cal L}=\{{\cal S}_\ell:\ell=1,\dots 5\}$ associated with the
potential, there is always a $6$th second order symmetry ${\cal S}_6$ that is
functionally dependent on $\cal L$, but linearly independent.
\end{corollary}

\section{Third order constants of the motion}\label{standardbasis}
The key to understanding the structure of the space of constants 
of the motion for superintegrable systems with nondegenerate potential is an  investigation of  third order constants 
of the motion. We have
\[
{\cal K}=\sum ^3_{k,j,i=1}a^{kji}(x,y,z)p_kp_jp_i+b^\ell(x,y,z)p_\ell,
\]
which must satisfy   
$\{{\cal H},{\cal K}\}=0$.
Here $a^{kji}$ is symmetric in the indices $k$, $j$, $i$.

The conditions are 
\begin{gather*}
a^{iii}_i=-\frac32 \sum_sa^{sii}(\ln\lambda)_s,\\
3a^{jii}_i+a^{iii}_j=-3\sum_sa^{sij}
(\ln\lambda)_s,\qquad i\ne j\nonumber\\
 a^{ijj}_i+a^{iij}_j=-\frac12 \sum_sa^{sjj}(\ln\lambda)_s-\frac12 \sum_sa^{sii}
(\ln\lambda)_s,\qquad i\ne j,\nonumber\\
2a^{ijk}_i+a^{kii}_j+ a^{jii}_k=- \sum_sa^{sjk}
(\ln\lambda)_s,\qquad i,\;j,\;k\ {\rm distinct},\nonumber\\
 b^j_k+ b^k_j   = 3\lambda \sum_sa^{skj}
V_s,\qquad j\ne k,\quad j,k=1,2,3,
\\
 b^j_j=\frac32 \lambda\sum_sa^{sjj}V_s-\frac12\sum_sb^s
(\ln\lambda)_s, \qquad j=1,2,3, 
\end{gather*}
and
\[ 
\sum_sb^{s}V_s=0.
\]
The $a^{kji}$ is just a third order Killing tensor. We are interested in such third order symmetries that could possibly arise as commutators of second order symmetries.
Thus we require that the highest order terms, the $a^{kji}$ in the constant of the motion, be independent of the four independent parameters in $V$. 
However, the $b^\ell$ must depend on these parameters.
We set
\[
b^\ell(x,y,z)=\sum_{j=1}^3f^{\ell,j}(x,y,z)V_j(x,y,z).
\]
(Here we are excluding the purely first order symmetries.) In
\cite{KKM20051} the following result is obtained. 

 \begin{theorem} Let $\cal K$ be a third order constant of the motion for a
  conformally flat superintegrable 
system with nondegenerate potential $V$:
\[
{\cal K}=\sum ^3_{k,j,i=1}a^{kji}(x,y,z)p_kp_jp_i+\sum ^3_{\ell=1}b^\ell(x,y,z)p_\ell.
\]
Then
\[
b^\ell(x,y,z)=\sum_{j=1}^3f^{\ell,j}(x,y,z) V_j(x,y,z)
\]
with 
$ f^{\ell,j}+f^{j,\ell}=0$, $1\le \ell,j\le 3$.
The $a^{ijk}$, $b^\ell$ are uniquely determined by the four
numbers 
\[
f^{1,2}(x_0,y_0,z_0),\quad f^{1,3}(x_0,y_0,z_0),\quad
f^{2,3}(x_0,y_0,z_0),\quad f^{1,2}_3(x_0,y_0,z_0) 
\]
at any regular point $(x_0,y_0,z_0)$ of $V$.
\end{theorem}

Let 
\[ {\cal S}_1=\sum a^{kj}_{(1)}p_kp_j+W_{(1)},\qquad  {\cal S}_2=\sum
a^{kj}_{(2)}p_kp_j+W_{(2)}
\]
be second order constants of the the motion for a superintegrable
system with
nondegenerate potential and let ${\cal A}_{(i)}(x,y,z)=\big\{a^{kj}_{(i)}(x,y,z)\big\}$, $i=1,2$ 
be $3\times 3$ matrix functions. Then the Poisson bracket of these
symmetries is given by
\[
\{{\cal S}_1,{\cal S}_2\}=\sum ^3_{k,j,i=1}a^{kji}(x,y,z)p_kp_jp_i+b^\ell(x,y,z)p_\ell,
\]
where
\[ 
f^{k,\ell}=2\lambda\sum_j\big(a^{kj}_{(2)}a^{j\ell}_{(1)}-a^{kj}_{(1)}a^{j\ell}_{(2)}\big).
\]
Differentiating, we find
\begin{gather} \label{commutatorcond1c}
f^{k,\ell}_i=2\lambda \sum_j
\big(\partial_ia^{kj}_{(2)}a^{j\ell}_{(1)}+a^{kj}_{(2)}\partial_ia^{j\ell}_{(1)} 
-\partial_ia^{kj}_{(1)}a^{j\ell}_{(2)} -a^{kj}_{(1)}\partial_ia^{j\ell}_{(2)}    \big )+G_if^{k,\ell}.
\end{gather}

Clearly,
$\{{\cal S}_1,{\cal S}_2\}$
is uniquely determined by the skew-symmetric matrix 
\[[{\cal A}_{(2)},{\cal A}_{(1)}]\equiv {\cal A}_{(2)}
{\cal A}_{(1)}-{\cal A}_{(1)}{\cal A}_{(2)}
,\]
hence by the constant matrix  $[{\cal A}_{(2)}(x_0,y_0,z_0),{\cal
  A}_{(1)}(x_0,y_0,z_0)]$ evaluated at a regular point, and by the number ${\cal F}(x_0,y_0,z_))=f^{1,2}_3(x_0,y_0,z_0)$.

For superintegrable nondegenerate potentials there is a standard
structure allowing the identification of the space of  second order constants of the
motion with the space $S_3$ of $3\times 3$ symmetric matrices, as well as
identification of the space of  third order constants of the
motion with a subspace of the space $K_3\times F$ of $3\times 3$ skew-symmetric matrices $K_3$
crossed with the line $F=\{{\cal F}(\boldsymbol{x}_0)\}$. Indeed,
if $\boldsymbol{x}_0$ is a regular point then there is a $1-1$ linear
correspondence between second order symmetries $\cal S$ and their associated
symmetric matrices ${\cal A}(\boldsymbol{x}_0)$. Let
$\{{\cal S}_1,{\cal S}_2\}'=\{{\cal S}_2,{\cal S}_1\}$ be the reversed Poisson bracket. Then the
map 
\[ \{{\cal S}_1,{\cal S}_2\}'\Longleftrightarrow  [{\cal A}_{(1)}(\boldsymbol{x}_0),{\cal A}_{(2)}(\boldsymbol{x}_0)]
\]
is an algebraic homomorphism. Here, ${\cal S}_1$, ${\cal S}_2$ are in involution if and
only if matrices ${\cal A}_{(1)}(\boldsymbol{x}_0)$, ${\cal A}_{(2)}(\boldsymbol{x}_0)$ 
commute and ${\cal F}(\boldsymbol{x}_0)=0$. If $\{{\cal S}_1,{\cal S}_2\}\ne 0$ then it is a third order
symmetry and can be uniquely associated with the skew-symmetric 
matrix  $[{\cal A}_{(1)}(\boldsymbol{x}_0),{\cal A}_{(2)}(\boldsymbol{x}_0)]$ 
and the parameter~${\cal F}(\boldsymbol{x}_0)$ . Let ${\cal E}^{ij}$ be the $3\times
3$ matrix with a $1$ in row $i$, column $j$ and $0$ for every other
matrix element. Then the  matrices
\begin{gather} \label{symbasisc} 
{\cal A}^{(ij)}=\frac12({\cal E}^{ij}+{\cal E}^{ji})={\cal
  A}^{(ji)},\qquad i,j=1,2,3
\end{gather}
form a basis for the $6$-dimensional space of symmetric
matrices. Moreover,
\[ 
[{\cal A}^{(ij)},{\cal A}^{(k\ell)}]=\frac12\big(
\delta_{jk}{\cal B}^{(i\ell)}+\delta_{j\ell}{\cal B}^{(ik)}+
\delta_{ik}{\cal B}^{(j\ell)}+\delta_{i\ell}{\cal B}^{(jk)}\big),
\]
where
\[
 {\cal B}^{(ij)}=\frac12({\cal E}^{ij}-{\cal E}^{ji})=-{\cal
  B}^{(ji)},\qquad i,j=1,2,3.
\]
Here ${\cal B}^{(ii)}=0$ and ${\cal B}^{(12)}$, ${\cal B}^{(23)}$, ${\cal
  B}^{(31)}$
form a basis for the space of skew-symmetric matrices. 
To obtain the  commutation relations for the second order
symmetries we need to use relations (\ref{commutatorcond1c}) 
to compute the parameter ${\cal F}(\boldsymbol{x}_0)$ associated with each commutator 
$[{\cal A}^{(ij)},{\cal A}^{(k\ell)}]$. 
The results are straightforward to compute, using relations (\ref{symmetryeqnsc}).

\[ 
\begin{array}{l@{\qquad }l} {\rm Commutator} &3{\cal F}/\lambda\\
\hline
\\[-3mm]
{}[{\cal A}^{(12)},{\cal A}^{(11)}]={\cal B}^{(21)} & -3 A^{13}- B^{23}-G_3\\[1mm]
{}[{\cal A}^{(13)},{\cal A}^{(11)}]={\cal B}^{(31)} &A^{12} -B^{33}+G_2\\[1mm]
{}[{\cal A}^{(22)},{\cal A}^{(11)}]=0 & -4A^{23}\\[1mm]
{}[{\cal A}^{(23)},{\cal A}^{(11)}]=0 & 2(A^{22}-A^{33})\\[1mm]
{}[{\cal A}^{(33)},{\cal A}^{(11)}]=0 &4A^{23}\\[1mm]
{}[{\cal A}^{(13)},{\cal A}^{(12)}]= \frac12{\cal B}^{(32)}&\frac12(3 B^{12}- A^{22}+3 A^{33}-G_1)\\[1mm]
{}[{\cal A}^{(22)},{\cal A}^{(12)}]= {\cal B}^{(21)}&-3B^{23}-A^{13}-G_3\\[1mm]
{}[{\cal A}^{(23)},{\cal A}^{(12)}]= \frac12{\cal B}^{(31)}&\frac12( -3B^{33}-3 A^{12}+2 B^{22}+G_2)\\[1mm]
{}[{\cal A}^{(33)},{\cal A}^{(12)}]= 0& 2(B^{23}-A^{13})\\[1mm]
{}[{\cal A}^{(22)},{\cal A}^{(13)}]= 0& -2B^{33}\\[1mm]
{}[{\cal A}^{(23)},{\cal A}^{(13)}]= \frac12{\cal B}^{(21)}& -
C^{33}+\frac12 B^{23}-\frac12 A^{13}-\frac12 G_3\\[1mm]
{}[{\cal A}^{(33)},{\cal A}^{(13)}]= {\cal B}^{(31)}&  A^{12}+B^{33}+G_2\\[1mm]
{}[{\cal A}^{(23)},{\cal A}^{(22)}]= {\cal B}^{(32)}&A^{33}-  A^{22}-B^{12}-G_1\\[1mm]
{}[{\cal A}^{(33)},{\cal A}^{(22)}]= 0&-4A^{23}\\[1mm]
{}[{\cal A}^{(33)},{\cal A}^{(23)}]= {\cal B}^{(32)}&A^{22}-A^{33}-B^{12}-G_1\\[1mm] 
\hline
\end{array}
\] 

A consequence of these results is \cite{KKM20051}

\begin{corollary} Let $V$ be a superintegrable nondegenerate
  potential on a conformally flat space, not a St\"ackel transform of the isotropic oscillator. Then
  the space of truly third order constants of the motion is $4$-dimensional
  and is spanned by Poisson brackets of the second order constants of
  the motion.
\end{corollary}

\begin{corollary} 
We can define a standard set of $6$ second order basis symmetries 
\[ {\cal S}^{(jk)}=\sum a_{(jk)}^{hs}(\boldsymbol{x})p_hp_s+W^{(jk)}(\boldsymbol{x})
\]
 corresponding to a regular point $\boldsymbol{x}_0$ by 
$(a_{(jk)})(\boldsymbol{x}_0)=A^{(jk)}$, $W^{(jk)}(\boldsymbol{x}_0)=0$.
\end{corollary}

\section{Maximum dimensions of the spaces of polynomial constants}\label{section5}

In order to demonstrate the existence and structure of quadratic algebras for 3D nondegenerate superintegrable systems
on conformally flat spaces, it is important to compute the dimensions of the spaces of 
symmetries of these systems that are of orders  4 and 6.   These symmetries are necessarily of a special type. 
The highest order terms in the momenta are independent
of the parameters in the potential, while 
the terms of order 2 less in the momenta are linear in these parameters, those of 
order 4 less are quadratic, and those of order 6 less are cubic. 
We will obtain these dimensions exactly, but first we need to establish sharp upper bounds. 

The following results are obtained by a careful study of the defining conditions 
and the integrability conditions for higher order constants of the motion~\cite{KKM20051}:
 \begin{theorem}
The maximum
possible dimension of the space of purely fourth order symmetries for
a nondegenerate {\rm 3D} potential is $21$.
The maximal possible dimension of the space of 
truly  sixth order symmetries is  $56$.
\end{theorem}

\section{Bases for the fourth and sixth order constants of the motion}

It follows from Section \ref{section5} that, for a superintegrable
system with nondegenerate potential, the dimension of the space of
truly fourth order constants of the motion is at most 21. Note from
Section \ref{standardbasis} that at
any regular point $\boldsymbol{x}_0$, we can define a standard basis of 6
second order constants of the motion ${\cal S}^{(ij)}=
  A^{(ij)}+W^{(ij)}$
where the quadratic form $A^{(ij)}$ has matrix ${\cal A}^{(ij)}$  defined by~(\ref{symbasisc}) 
and $W^{(ij)}$ is the potential term
with $W^{(ij)}(\boldsymbol{x}_0)\equiv 0$ identically in the parameters
$W^{(\alpha)}$. By taking homogeneous polynomials of order two in the
standard basis symmetries we can construct fourth order symmetries. 

\begin{theorem}\label{fourthorderbasis} The $21$ distinct standard monomials 
$ {\cal S}^{(ij)}{\cal S}^{(jk)}$, defined with respect to a regular
point $\boldsymbol{x}_0$, form a basis for the space of fourth
order symmetries.
\end{theorem}

Indeed, we can choose the basis symmetries in the form
\[ 
1. \ \ \big({\cal S}^{(ii)}\big)^2,\ \ {\cal S}^{(ii)}{\cal
    S}^{(ij)},\ \ {\cal S}^{(ii)}{\cal
    S}^{(jj)},\ \ {\cal S}^{(ii)}{\cal
    S}^{(jk)}
\]
for $i,j,k=1,\dots,3$, $i$, $j$, $k$ pairwise distinct (15 possibilities).
\[
2.\ \ {\cal S}^{(ii)}{\cal
    S}^{(jj)}-\big({\cal S}^{(ij)}\big)^2
\]
for $i,j=1,\dots,3$, $i$, $j$ pairwise distinct (3 possibilities).
\[
3. \ \ {\cal S}^{(ij)}{\cal
    S}^{(ik)}-{\cal S}^{(ii)}{\cal S}^{(jk)}
\]
for $i,j,k=1,\dots,3$, $i$, $j$, $k$ pairwise distinct (3 possibilities).

It is a straightforward computation to show that these 21 symmetries are linearly independent. 
Since the maximum possible dimension of the space of fourth order symmetries is 21, they must form a basis. 
See \cite{KKM20051} for the details of the proof.

Now  from Section \ref{section5} the dimension of the
space of purely sixth order constants of the motion is at most
56. Again we can show that the 56 independent homogeneous third order polynomials
in the symmetries ${\cal
    S}^{(ij)}$ form a basis for this space.

At the sixth order level we have the symmetries 
\[ 
1. \ \ \big({\cal S}^{(ii)}\big)^3,\ \  \big({\cal S}^{(ii)}\big)^2{\cal
    S}^{(ij)},\ \ \big({\cal S}^{(ii)}\big)^2{\cal
    S}^{(jj)},\  \ \big({\cal S}^{(ii)}\big)^2{\cal
    S}^{(jk)}
\]
for $i,j,k=1,\dots,3$, $i$, $j$, $k$ pairwise distinct (18 possibilities).
\[
2. \ \ {\cal S}^{(ii)}{\cal S}^{(ij)}{\cal S}^{(jj)},\ \ {\cal S}^{(ii)}{\cal
    S}^{(ij)}{\cal     S}^{(jk)},\  \ {\cal S}^{(ii)}{\cal
    S}^{(jj)}{\cal S}^{(kk)},
\]
for $i,j,k=1,\dots,3$, $i$, $j$, $k$ pairwise distinct (10 possibilities).
\[
3. \ \  {\cal S}^{(\ell m)}\big({\cal S}^{(ii)}{\cal
    S}^{(jj)}-\big({\cal S}^{(ij)}\big)^2\big)
\]
for $i,j=1,\dots,3$, $i$, $j$ pairwise distinct (10 possibilities).
\[
4. \ \ {\cal S}^{(\ell m)}\big({\cal S}^{(ij)}{\cal
    S}^{(ik)}-{\cal S}^{(ii)}{\cal S}^{(jk)}\big)
\]
for $i,j,k=1,\dots,3$, $i$, $j$, $k$ pairwise distinct (18  possibilities).

\begin{theorem} The $56$ distinct standard monomials  ${\cal
    S}^{(hi)}{\cal S}^{(jk)}{\cal S}^{(\ell m)}$, defined with respect
  to a~re\-gu\-lar $\boldsymbol{x}_0$, form a basis for the space of sixth order
  symmetries.
\end{theorem}

See \cite{KKM20051} for the details of the proof. 
 We conclude that the
quadratic algebra closes. 

\section{Second order conformal Killing tensors}\label{conformalKt}
There is a close relationship between the second-order Killing
tensors of a conformally flat space in 3D and the second order
conformal Killing tensors of flat space. A second order  conformal
Killing tensor for a space $\cal V$ with metric
$ds^2=\lambda(\boldsymbol{x})\big(dx_1^2+dx_2^2+dx_3^2\big)$ and free Hamiltonian
${\cal H}=\big(p_1^2+p_2^2+p_3^2\big)/\lambda$ is a quadratic form ${\cal
  S}=\sum a^{ij}(x_1,x_2,x_3)p_ip_j$ such that
\[
\{ {\cal H},{\cal S}\}=f(x_1,x_2,x_3){\cal H},
\]
form some function $f$. Since $f$ is arbitrary, it is easy to see that
$\cal S$ is a conformal Killing tensor for $\cal V$ if and only if it is
a conformal Killing tensor for flat space $dx_1^2+dx_2^2+dx_3^2$. The
conformal Killing tensors for flat space are very well known, e.g.,
\cite{ERNIE}.
The space of conformal Killing tensors is infinite dimensional.
It is spanned by products of the conformal Killing vectors
\begin{gather*}
 p_1,\ \  p_2,\ \ p_3,\  \ x_3p_2-x_2p_3,\ \ x_1p_3-x_3p_1,\ \ x_2p_1-x_1p_2,\ \
x_1p_1+x_2p_2+x_3p_3,
\\
\left(x_1^2-x_2^2-x_3^2\right)p_1+2x_1x_3p_3+2x_1x_2p_2,\ \ 
\left(x_2^2-x_1^2-x_3^2\right)p_2+2x_2x_3p_3+2x_2x_1p_1,\  \
\\
\left(x_3^2-x_1^2-x_2^2\right)p_3+2x_3x_1p_1+2x_3x_2p_2,
\end{gather*}
and terms $g(x_1,x_2,x_3)\big(p_1^2+p_2^2+p_3^2\big)$ where $g$ is an
arbitrary function. Since every Killing tensor is also a conformal
Killing tensor, we see that every second-order Killing tensor for
$V_3$ can be expressed as a linear combination of these second-order
generating elements though, of course, the space of Killing tensors is
only finite dimensional. This shows in particular that every $a^{ij}$
and every $a^{ii}-a^{jj}$
with $i\ne j$ is  a polynomial of order at most 4 in $x_1$, $x_2$, $x_3$, no
matter what is the choice of $\lambda$.

It is useful to pass to new variables 
\[
 a^{11},\ \ a^{24},\ \ a^{34},\ \ a^{12},\ \ a^{13},\ \ a^{23}
\]
for the Killing tensor, where $a^{24}=a^{22}-a^{11}$, $a^{34}=a^{33}-a^{11}$. 
Then we can establish the result
\begin{theorem} Necessary and sufficient conditions that the quadratic form 
${\cal S}=\sum\limits_{ij}a^{ij}p_ip_j$ be a~second order Killing tensor for 
the space with metric $ds^2=\lambda\big(dx_1^2+dx_2^2+dx_3^2\big)$
are 
\smallskip

$1.$~$\cal S$ is a conformal Killing tensor on the flat space with metric $dx_1^2+dx_2^2+dx_3^2$.

$2.$~The following integrability conditions  hold:
\begin{gather*}
 \left(\lambda_2a^{12}+\lambda_3a^{13}\right)_2
 = \left(\lambda_1a^{12}+\left(a^{24}\lambda\right)_2+\lambda_3a^{23}\right)_1,
\\
 \left(\lambda_2a^{12}+\lambda_3a^{13}\right)_3= \left(\lambda_1a^{13}+\lambda_2a^{23}    +
 \left(a^{34}\lambda\right)_3  \right)_1,
\\
 \left(\lambda_1a^{12}+\left(a^{24}\lambda\right)_2+\lambda_3a^{23}\right)_3
 =\left(\lambda_1a^{13}+\lambda_2a^{23}   
  +\left(a^{34}\lambda\right)_3  \right)_2.
\end{gather*}
\end{theorem}

\section{The St\"ackel transform for three-dimensional systems}
The St\"ackel transform \cite{BKM} or coupling constant metamorphosis \cite{HGDR} plays a fundamental role in relating superintegrable systems on different manifolds.
Suppose we have a superintegrable system 
\[  
H=\frac{p_1^2+p_2^2+p^2_3}{\lambda(x,y,z)}+V(x,y,z),
\]
in local orthogonal coordinates, with  nondegenerate potential $V(x,y,z)$:
\begin{gather}
V_{33}=V_{11}+A^{33}V_1+B^{33}V_2+C^{33}V_3,\nonumber\\
V_{22}=V_{11}+A^{22}V_1+B^{22}V_2+C^{22}V_3,\nonumber\\
V_{23}= \phantom{V_{11}+{}}A^{23}V_1+B^{23}V_2+C^{23}V_3,\nonumber\\
V_{13}= \phantom{V_{11}+{}}A^{13}V_1+B^{13}V_2+C^{13}V_3,\nonumber\\
V_{12}= \phantom{V_{11}+{}}A^{12}V_1+B^{12}V_2+C^{12}V_3\label{veqn21}
\end{gather}
and suppose $U(x,y,z) $ is a particular solution of equations (\ref{veqn21}), nonzero in an open set.  Then
the transformed system 
\[  
{\tilde H}=\frac{p_1^2+p_2^2+p^2_3}{{\tilde \lambda}(x,y,z)}+{\tilde V}(x,y,z)
\]
 with  nondegenerate potential ${\tilde V}(x,y,z)$:
\begin{gather*}
{\tilde V}_{33}={\tilde V}_{11}+{\tilde A}^{33}{\tilde V}_1+{\tilde B}^{33}{\tilde V}_2+{\tilde C}^{33}{\tilde V}_3,\\
{\tilde V}_{22}={\tilde V}_{11}+{\tilde A}^{22}{\tilde
  V}_1+{\tilde B}^{22}{\tilde V}_2+{\tilde C}^{22}{\tilde V}_3,\\
{\tilde V}_{23}=\phantom{{\tilde V}_{11}+{}}{\tilde A}^{23}{\tilde V}_1+{\tilde
  B}^{23}{\tilde V}_2+{\tilde C}^{23}{\tilde V}_3,\\
{\tilde V}_{13}=\phantom{{\tilde V}_{11}+{}}{\tilde A}^{13}{\tilde V}_1
+{\tilde B}^{13}{\tilde V}_2+{\tilde C}^{13}{\tilde V}_3,\\
{\tilde V}_{12}=\phantom{{\tilde V}_{11}+{}}{\tilde A}^{12}{\tilde V}_1
+{\tilde B}^{12}{\tilde V}_2+{\tilde C}^{12}{\tilde V}_3,
\end{gather*}
is also superintegrable, where
\begin{gather*}
{\tilde \lambda}=\lambda U,\qquad  {\tilde V}=\frac{V}{U},
\\
{\tilde
  A}^{33}=A^{33}+2\frac{U_1}{U},\qquad  {\tilde B}^{33}=B^{33}, \qquad {\tilde
C}^{33}=C^{33}-2\frac{U_3}{U},
\\
{\tilde
  A}^{22}=A^{22}+2\frac{U_1}{U},\qquad  {\tilde B}^{22}=B^{22}-2\frac{U_2}{U}, \qquad {\tilde
C}^{22}=C^{22},
\\
{\tilde
  A}^{23}=A^{23},\qquad  {\tilde B}^{23}=B^{23}-\frac{U_3}{U}, \qquad {\tilde
C}^{23}=C^{23}-\frac{U_2}{U},
\\
{\tilde
  A}^{13}=A^{13}-\frac{U_3}{U},\qquad  {\tilde B}^{13}=B^{13}, \qquad {\tilde
C}^{13}=C^{13}-\frac{U_1}{U},
\\
{\tilde
  A}^{12}=A^{12}-\frac{U_2}{U},\qquad  {\tilde B}^{12}=B^{12}-\frac{U_1}{U}, \qquad {\tilde
C}^{12}=C^{12}.
\end{gather*}
Let $S=\sum a^{ij}p_ip_j+W=S_0+W$  be a second order symmetry of $H$ 
and $S_U=\sum a^{ij}p_ip_j+W_U=S_0+W_U$  be the special case  that is in involution with 
$\big(p_1^2+p_2^2+p^2_3\big)/{ \lambda}+ U$. Then 
\[
{\tilde S}=S_0-\frac{W_U}{U}H+\frac{1}{U}H
\]
is the corresponding symmetry of $\tilde H$. Since one can always 
add a constant to a nondegenerate potential, it follows that $1/U$ 
defines an inverse St\"ackel transform of $\tilde H$ to $H$.
See \cite{BKM} for many examples of this transform.

\section{Multiseparability and St\"ackel equivalence}

{}From the general theory of variable separation for Hamilton--Jacobi
equations \cite{ERNIE, MIL88} we know that second order symmetries
${\cal S}_1$, ${\cal S}_2$
define a separable system for the equation
\[
 H=\frac{p^2_x+p^2_y+p^2_z}{\lambda(x,y,z)}+V(x,y,z)=E
\]
if and only if
\begin{enumerate}
\itemsep=-1pt
\item The symmetries ${\cal H}$, ${\cal S}_1$, ${\cal S}_2$ form a linearly independent set as
  quadratic forms.
\item $\{{\cal S}_1,{\cal S}_2\}=0$.
\item The three quadratic forms have a common eigenbasis of
  differential forms. 
\end{enumerate}
This last requirement means that, expressed in  coordinates $x$, $y$, $z$,
at least one of the matrices ${\cal A}_{(j)}(\boldsymbol{x})$ can be
diagonalized by conjugacy transforms in a neighborhood of a regular
point and that $[{\cal A}_{(2)}(\boldsymbol{x}),{\cal A}_{(1)}(\boldsymbol{x})]=0$.
However, for nondegenerate superintegrable potentials in a~conformally
flat space we see from  Section \ref{section5} that 
\[
 \{{\cal S}_1,{\cal S}_2\}=0 \ \Longleftrightarrow \ [{\cal A}_{(2)}(\boldsymbol{x}_0),{\cal
  A}_{(1)}(\boldsymbol{x}_0)]=0,\qquad {\rm and}\qquad  {\cal F}(\boldsymbol{x}_0)=0
\]
so that the intrinsic conditions for the existence of a separable
coordinate system are  simplified.

\begin{theorem}\label{3Dmultiseparable} Let $V$ be a superintegrable nondegenerate
  potential in a {\rm 3D} conformally flat  space. Then $V$ defines a
  multiseparable system.
\end{theorem}
See \cite{KKM20052} for the details of the proof. 

In \cite{KMR} the following result was obtained. 
\begin{theorem} \label{constantcurvsep} Let $u_1$, $u_2$, $u_3$ be an
orthogonal separable coordinate system  for a {\rm 3D}  conformally 
flat space with metric $d{\tilde s}^2$ Then there is a function $f$ such that 
\[
fd{\tilde s}^2= d{s}^2
\]
where $ds^2$ is a constant curvature space metric and $ds^2$ is
orthogonally separable in exactly these same coordinates
$u_1$, $u_2$, $u_3$. The function $f$ is called a St\"ackel multiplier with respect to this coordinate system. 
\end{theorem}
Thus the possible separable coordinate systems for a conformally 
flat space are all obtained, modulo a St\"ackel multiplier, 
from separable systems on 3D flat space or on the 3-sphere. This result provides evidence that, as in the 2D 
case, all nondegenerate 3D superintegrable systems on conformally flat 
spaces are St\"ackel equivalent to a superintegrable system on either 3D 
flat space or the 3-sphere, but we have not yet settled this issue.

\section{Discussion and conclusions}
We have shown that all classical superintegrable systems with
nondegenerate potential on real
or complex 3D conformally flat spaces admit 6 linearly independent
second order constants of the motion (even though only 5 functionally
independent second order constants are assumed) and that the spaces of
fourth order and sixth order symmetries are spanned by polynomials in
the second order symmetries. (An interesting issue here is the form of the functional
dependence relation between the 6 linearly independent symmetries. It
appears that the relation is always of order 8 in the momenta, but we
have as yet no general proof.) This implies that a quadratic algebra
structure always exists for such systems. We worked out their common
structure and related it to algebras of $3\times 3$ symmetric
matrices.  We demonstrated that such systems are
always multiseparable, more precisely  they permit separation of
variables in at least three orthogonal coordinate systems.

We also studied the St\"ackel transform, a conformal  invertible mapping from a
superintegrable system on one space to a system on another
space. Using prior results from the theory of separation of variables on conformally flat spaces, 
we gave evidence that every nondegenerate superintegrable system on such 
a space is St\"ackel equivalent to a superintegrable system on complex 
Euclidean space or on the complex 3-sphere, though we haven't yet 
settled the issue. This suggests that  to classify all such superintegrable systems we can restrict
attention to these two constant curvature spaces, and then obtain all
other cases via St\"ackel transforms. We are making considerable
progress on the classification theory \cite{KKM20052}, though the problem is
complicated.

All of our 2D and 3D classical results can be extended to quantum
systems and the Schr\"odinger equation and we are in the process of
writing these up. 

Another interesting set of issues comes from the consideration of 3D
superintegrable systems with degenerate, but multiparameter,
potentials. In some cases such as the extended Kepler--Coulomb potential
there is no quadratic algebra, whereas in other cases the quadratic
algebra exists. Understanding the underlying structure of these
systems is a major challenge. Finally there is the challenge of
generalizing the 2D and 3D results to higher dimensions.

\LastPageEnding
\end{document}